 \documentclass[aps,prl,reprint,superscriptaddress,amssymb,amsmath]{revtex4-1}
  \usepackage[utf8]{inputenc}
  \usepackage[T1]{fontenc}
  \usepackage[english]{babel}
  \usepackage{times}
  \usepackage{amssymb,amsmath,amsfonts,amsthm}
  \usepackage{mathtools}
  \usepackage{verbatim}
  \usepackage{enumerate}
  \usepackage{graphicx}
  \usepackage{hyperref}
  \usepackage{bbm}
  \usepackage{booktabs}
  
  \usepackage[caption=false]{subfig}
  \usepackage{mathrsfs}
  \usepackage{color}
  \usepackage{bm}
  \usepackage{braket}

\usepackage{graphicx}
\usepackage{multirow}
\usepackage{bm}
\begin{document}

\title{Measurement-Device-Independent Quantum Secure Direct Communication}
\author{Zeng-Rong Zhou}
\thanks{These authors made equal contribitutions}
\affiliation{State Key Laboratory of Low-dimensional Quantum Physics, Beijing, 100084, China}
\affiliation{Department of Physics, Tsinghua University, Beijing, 100084, China}
\affiliation{Beijing National Research Center for Information Science and Technology, Beijing, 100084, China}
\affiliation{School of Information and Technology,Tsinghua University, Beijing, 100084, China}
\affiliation{Collaborative Innovation Center of Quantum Matter, Beijing, 100084, China}
\affiliation{Beijing Academy of Quantum Information, Beijing, 100084, China}

\author{Yu-Bo Sheng}
\thanks{These authors made equal contribitutions}
\affiliation{Institute of Quantum Information and Technology, Nanjing University of Posts and Telecommunications, Nanjing, 210003, China}
\affiliation{College of Telecommunications \& Information Engineering, Nanjing University of Posts and Telecommunications, Nanjing, 210003, China}
\affiliation{Key Lab of Broadband Wireless Communication and Sensor Network Technology, Nanjing University of Posts and Telecommunications, Ministry of Education, Nanjing, 210003, China}

\author{Peng-Hao Niu}
\affiliation{State Key Laboratory of Low-dimensional Quantum Physics, Beijing, 100084, China}
\affiliation{Department of Physics, Tsinghua University, Beijing, 100084, China}
\affiliation{Beijing National Research Center for Information Science and Technology, Beijing, 100084, China}
\affiliation{School of Information and Technology,Tsinghua University, Beijing, 100084, China}
\affiliation{Collaborative Innovation Center of Quantum Matter, Beijing, 100084, China}
\affiliation{Beijing Academy of Quantum Information, Beijing, 100084, China}

\author{Liu-Guo Yin}
\thanks{Corresponding author: yinlg@tsinghua.edu.cn},
\affiliation{Beijing National Research Center for Information Science and Technology, Beijing, 100084, China}
\affiliation{School of Information and Technology,Tsinghua University, Beijing, 100084, China}

\author{Gui-Lu Long}
\thanks{Corresponding author: gllong@tsinghua.edu.cn}
\affiliation{State Key Laboratory of Low-dimensional Quantum Physics, Beijing, 100084, China}
\affiliation{Department of Physics, Tsinghua University, Beijing, 100084, China}
\affiliation{Beijing National Research Center for Information Science and Technology, Beijing, 100084, China}
\affiliation{School of Information and Technology,Tsinghua University, Beijing, 100084, China}
\affiliation{Collaborative Innovation Center of Quantum Matter, Beijing, 100084, China}
\affiliation{Beijing Academy of Quantum Information, Beijing, 100084, China}

\begin{abstract}
 Quantum secure direct communication (QSDC) is the technology to transmit secret information directly through a quantum channel without neither key nor ciphertext. It provides us with a secure communication structure that is fundamentally different from the one that we use today. In this Letter, we report the first measurement-device-independent(MDI) QSDC protocol with sequences of entangled photon pairs and single photons. It eliminates security loopholes associated with the measurement device. In addition, the MDI technique doubles the communication distance compared to those without using the technique.  We also give a protocol with linear optical Bell-basis measurement, where only two of the four Bell-basis states could be measured.  When the number of qubit in a sequence reduces to 1, the MDI-QSDC protocol reduces to a deterministic MDI quantum key distribution protocol, which is also presented in the Letter.
\end{abstract}

\maketitle

\paragraph{Introduction}

\begin{figure}
  \begin{center}
  \includegraphics[width=0.5\textwidth]{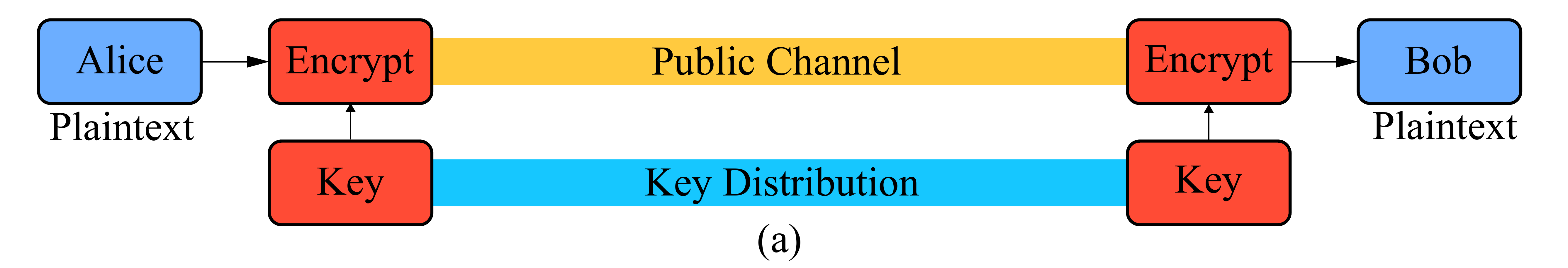}
  \includegraphics[width=0.5\textwidth]{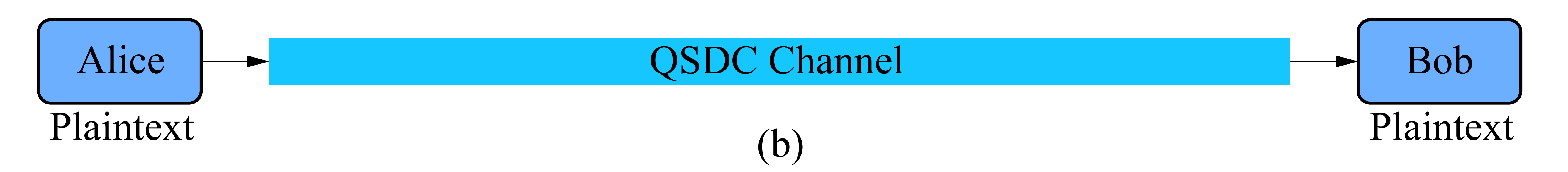}
  \caption{(coloronline)(a) Structure of a general secure communication. Usually key distribution is completed using RSA. With QKD, unconditional key distribution can be achieved. (b) The structure of quantum secure direct communication. It has no key distribution, key storage and management, and no ciphertext. It eliminates all three security loopholes in a general secure communication structure.}\label{fig1-secure}
  \end{center}
  \end{figure}

A secure communication structure is usually composed of a key distribution channel and a ciphertext transmission channel, as shown in Fig. \ref{fig1-secure}a. Usually, ciphertext is encoded with the AES cipher \cite{aes}, and key is distributed using RSA public cryptosystem \cite{rsa}.  There are three potential security loopholes in this structure: leak of key during distribution, loss of key in storage and in transition at users' sites, and interception of ciphertext in transmission.  Quantum principle enables legitimate users to detect Eve,  and quantum  key distribution (QKD) offers provably security for key agreement \cite{Bennett:1984wva,Ekert:1991kl}. In QKD, Alice sends random numbers encoded in quantum states to Bob. They can detect Eve by publicly comparing some samples. If Eve is found, they discard the transmitted data because all or part of them has already leaked. If they are certain that there is no eavesdropping, the transmitted data will be used as key to encode a message into ciphertext.   QKD eliminates in principle the security loophole in the key distribution channel, but the other two loopholes still persist. An Eve can always intercept the ciphertext and store them for cryptanalysis. Only Vernam's one-time-pad was shown to be perfectly secure \cite{vernam,shannon} provided the key is absolutely protected. Though happened very rarely, loss of key or repeated use of key still occurred, with disastrous consequences \cite{camfive}.

 QSDC \cite{[][{, also at Epreprint arXiv:quant-ph/0012056. }]Long:2002ei, Deng:2003ih,Deng:2004kg} transmits a message directly over a quantum channel. It does not use key, hence there is no key distribution and key storage and management. The security loopholes associated with the key are all eliminated. QSDC establishes a secure quantum channel first, and any attempt to intercept the QSDC channel would obtain only random numbers, hence the security loophole associated with the ciphertext is also wiped out. As shown in Fig.\ref{fig1-secure}b,  QSDC eliminates all three security loopholes in traditional secure communication, and changes fundamentally the structure. This could lead further changes in future secure communication.

Recently, there have been remarkable developments in experimental QSDC. A single-photon QSDC protocol with error correction code was proposed  and experimentally demonstrated \cite{Hu:2016gz}. It has shown that QSDC works in noisy and lossy environment.    QSDC protocols based on Einstein-Podolsky-Rosen (EPR) pair in Refs.\cite{Long:2002ei,Deng:2003ih} have been experimentally realized using atomic quantum memory in optical platform \cite{Zhang:2017ew}, and fiber-photonics devices at a  distance of a few kilo-meters \cite{Zhu:2017ie} respectively. They have attracted widespread attention both in the academic and security circles \cite{circle2017}.

 All components in a practical setting have defects and imperfections, and they can be used to steal  secret key in practical QKD systems \cite{Fung:2007hf,Xu:2010ka,Zhao:2008jn,Lydersen:2010jm,light}. Among these loopholes, those in the measurement devices are dominant. One solution to this problem is to fabricate near perfect devices.  However it takes somehow long time to advance related fabrication technology, and there is  always some degree of inaccuracy in any real device. An alternative way is to design new protocols, taking into account the imperfections existent in these devices. This has been successfully done in QKD in the measurement-device-independent(MDI) protocol  \cite{Lo:2012dfa}. In the MDI technique, the measurement-device is in principle in the hands of an untrusted Charlie who performs the measurement. It eliminates the security loopholes in the measurement part of the system.

QSDC is secure under ideal conditions, such as perfect quantum source, noiseless channel, perfect devices and detectors \cite{Long:2002ei,Deng:2003ih,Deng:2004kg,Li:2015dv,Lu:2011hl}. In order for QSDC to go for practical application, measurement-device-independent QSDC is essential.
 In this Letter, we propose a measurement-device-independent QSDC (MDI-QSDC) protocol based on single photons and entangled photon pairs. In this scheme, Alice prepares a sequence of EPR-pairs, and  Bob prepares a sequence of single photon states and "send" them  to Alice through teleportation, in which the Bell-basis measurement is performed by an untrusted Charlie. Then after checking the security, Alice encodes her message in the teleported single photons, and sends them to Charlie who performs the single qubit measurement for Bob. MDI-QSDC enhances greatly the security of QSDC under realistic condition. In addition to the security advantage, MDI-QSDC effectively doubles the communication distance because both Alice and Bob send their qubits to the measurement-device which lies in the middle of them. We also give a protocol with linear optics Bell-basis measurement, where only two of the four Bell-basis states could be measured. QSDC enables the direct secure transmission of information by the block data transmission technique \cite{Long:2002ei,Deng:2003ih}, in which the quantum information carriers are transmitted in a block of large number of qubits. When the number of qubit in a block is reduced to 1, the MDI-QSDC protocol is reduced to a deterministic MDI-QKD protocol, which is also described toward the end of the Letter.

\begin{figure}
  \begin{center}
  \includegraphics[width=0.3\textwidth]{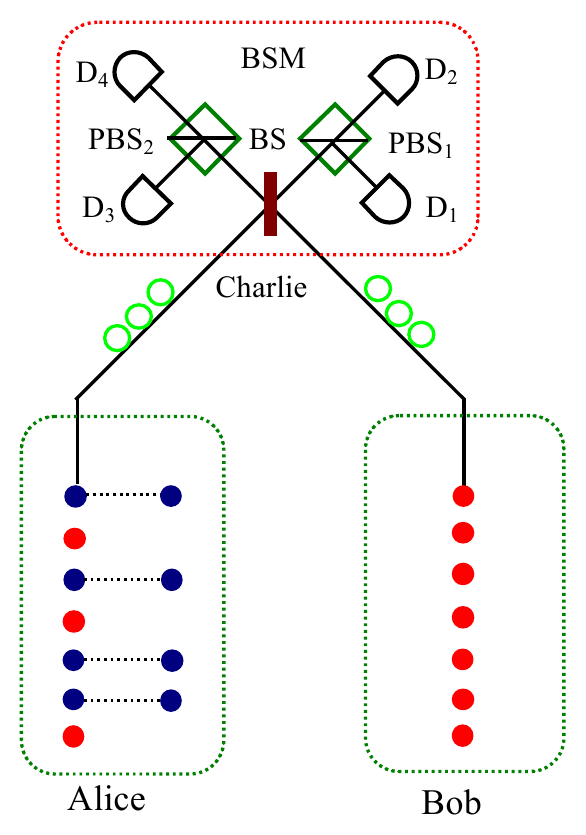}
  \caption{(coloronline)Sketch for experimental implementation of MDI-QSDC}\label{fig2}
  \end{center}
  \end{figure}

\paragraph{Method}
  The protocol uses both  Bell-basis states,
  \begin{equation}
| \phi^\pm \rangle =  \left(|00\rangle \pm |11\rangle \right)/\sqrt{2},
| \psi^\pm \rangle =  \left(|01\rangle \pm |10\rangle \right)/\sqrt{2},
\label{eq:bell}
\end{equation}
and single qubit states  $\ket{\pm}=(\ket{0}\pm\ket{1})/\sqrt{2}$,  $\ket{0}$ and $\ket{1}$.
  The protocol consists of the following 6 steps, and we suppose Alice sends information to Bob. The protocol is illustrated in Fig.\ref{fig2}.

  Step 1) Alice and Bob prepare ordered qubits sequences.   Alice produces a sequence of $N+t_0$ EPR-pairs in Bell-state $\ket{\psi^{-}_{12}}$ in her site. She divides her EPR pair sequence into two single qubit sequences, $S_{Ah}$ and $S_{At}$, whose qubits are partners each other in the EPR pairs. She also prepares a sequence of $t_1$ number of single qubits whose states are randomly in  one of the  $\ket{+}$, $\ket{-}$, $\ket{0}$ states, $\ket{1}$, and inserts them into $S_{At}$ in random positions so as to form an ordered sequence $P_{A}$ of $N+t_0+t_1$ single qubits.  Meanwhile, Bob prepares a sequence of $N+t_0+t_1$ single qubits, $P_B$, whose states are randomly in  one of the four states $\ket{+}$, $\ket{-}$, $\ket{0}$ and $\ket{1}$.

 The EPR pairs in Alice's side are used for directly communicating secret information, and the single qubits are used for security check.

  Step 2). Alice sends sequence $P_{A}$, and Bob sends sequence $P_B$ to Charlie. Charlie performs Bell-basis measurement on every pair of qubits he receives from Alice and Bob, and publishes the results.  The Bell-basis measurement of a single qubit from an EPR-pair of $P_A$ with a single qubit from $P_B$ leads to the collapse of Alice's EPR-pair into one of the four single qubit states, $\{\ket{+}$, $\ket{-}$, $\ket{0}$, $\ket{1}\}$ with equal probabilities, as shown in Table \ref{t1}. The state after measurement is only known to Bob, and unknown to  both Alice and Eve. This is a quantum teleportation process in a slightly complicated manner, where Bob's single qubit is almost teleported Alice, apart from a unitary operation $U_T$ to transform Alice's corresponding qubit in $S_{Ah}$ into Bob's state. $U_T$ is known to all Alice, Bob and Charlie after Charlie announces his Bell-basis measurement result.  For instance, for $\ket{\psi^-_{12}}\ket{0}_3$, $U_T=I$ if the Bell-basis measurement yields $\ket{\psi_{23}^+}$   or $\ket{\psi_{23}^-}$; $U_T=i\sigma_Y$, if the Bell-basis measurement yields $\ket{\phi_{23}^+}$ or  $\ket{\phi_{23}^-}$.
     \begin{table}
  \begin{ruledtabular}
  \begin{tabular}{l|rrrr}\hline
  Bob's state      & $\ket{\phi_{23}^+}$ & $\ket{\phi_{23}^-}$ & $\ket{\psi_{23}^+}$      & $\ket{\psi_{23}^-}$ \\ \hline
  $\ket{0}_3$ &       $-\ket{1}_1$          &  $-\ket{1}_1$             & $-\ket{0}_1$               & $-\ket{0}_1$ \\
  $\ket{1}_3$ &       $\ket{0}_1$           &  $-\ket{0}_1$             & $-\ket{1}_1$              & $-\ket{1}_1$ \\
  $\ket{+}_3$ &       $\ket{-}_1$           &  $-\ket{+}_1$             & $\ket{-}_1$               & $-\ket{+}_1$ \\
  $\ket{-}_3$ &       $-\ket{+}_1$          &  $\ket{-}_1$              & $\ket{+}_1$               & $-\ket{-}_1$\\ \hline
  \end{tabular}
  \end{ruledtabular}
  \caption{$\ket{\psi^-}_{12}\ket{q}_3$ in terms of Bell-states of qubit 2 and 3. In front of each term, there is a coefficient $1/2$. }\label{t1}
  \end{table}

Step 3). Security check. Alice publishes the positions and states of the $t_1$ single qubits in $P_A$, and Bob also publishes the corresponding states of the corresponding qubits in $P_B$.  This security check is identical to that in the MDI-QKD.
 For those qubits whose basis are different,  a Bell-basis measurement will yield any one of the four Bell-basis states, as shown in Eq. (\ref{ES8}),
   \begin{eqnarray}
   \ket{+}\ket{0}&=+\frac{1}{2}\ket{\phi^+}+\frac{1}{2}\ket{\phi^-}+\frac{1}{2}\ket{\psi^+}+\frac{1}{2}\ket{\psi^-},\nonumber\\
   \label{ES8}\ket{-}\ket{1}&=-\frac{1}{2}\ket{\phi^+}+\frac{1}{2}\ket{\phi^-}+\frac{1}{2}\ket{\psi^+}+\frac{1}{2}\ket{\psi^-}.
   \end{eqnarray}
There are not useful for security check. The decomposition of  qubits with identical basis  in terms of Bell-basis states are shown in Eq. (\ref{ES5}).
  \begin{eqnarray}
  \ket{+}\ket{+}=\frac{1}{\sqrt{2}}\left(\ket{\phi^+}+\ket{\psi^+}\right), &\ket{+}\ket{-}=\frac{1}{\sqrt{2}}\left(\ket{\phi^-}-\ket{\psi^-}\right),\nonumber \\
  \ket{0}\ket{0}=\frac{1}{\sqrt{2}}\left(\ket{\phi^+}+\ket{\phi^-}\right), &\ket{0}\ket{1}=\frac{1}{\sqrt{2}}\left(\ket{\psi^+}+\ket{\psi^-}\right).\nonumber \\ \label{ES5}
  \end{eqnarray}
    A Bell-basis measurement can only obtain one of two Bell-basis states. Charlie's eavesdropping will have a 50\% probability to obtain the other two Bell-basis states, hence increases the error rate. If the error rate is above the threshold, then the process will be terminated. Otherwise, go to the next step.

 Step 4). Encoding message by Alice. Bob announces the basis of his remaining qubits.   Alice's encoding operation $U$ is the product of two operations, $U=U_mU_T$. The first one is the unitary operation to complete the quantum teleportation $U_T$, and the other operation is the message encoding operation $U_m$, namely $I$ for 0, which does not change the state at all, and  $i\sigma_Y$ for 1, which flips the state. To ensure the integrity of the message, Alice also encodes $t_0$ qubits with random numbers, which are positioned randomly in $S_{Ah}$.

 Step 5). Alice sends the sequence to Charlie. Charlie first performs unitary operations, $U_B$s,  so that the qubit basis becomes $\{$ $\ket{0}$, $\ket{1}$ $\}$ , namely $U_B=I$ if the basis of Bob's qubit is $\{\ket{0}$, $\ket{1}\}$, and $U_B=H$  if  the basis of Bob's qubit is $\{\ket{+}$, $\ket{-}\}$.  Charlie measures the qubits in sequence $S_{Ah}$ in $\sigma_Z$ basis and announces the results. Upon these results, Bob can derive the message and random numbers encoded by Alice.

 Step 6). Integrity check. Alice announces the random numbers of the $t_0$ check qubits. If the error rate is below threshold, the transmission is safe. Alice and Bob conclude that the direct communication is secure and complete the session. Otherwise, they conclude the communication is tampered by Eve or the untrusted Charlie. It should be emphasized that Eve or Charlie's eavesdropping at step 5) could not steal any information, it can only disturb the communication. Integrity check ensures the message Bob receives is correct.

 \paragraph{Security analysis}  The security check in step 3) is identical to that in MDI-QKD. After the security check, the security of teleportation process of Bob's qubits is ascertained. Before the Bell-basis measurement, the density matrix of qubit in Alice's  sequence $P_A$ is composed qubits from $N+t_0$ entangled  EPR-pairs whose density is ${\rm Tr}_1(\ket{\psi^-_{12}}\langle \psi^-_{12}|)$ and a sequence of $t_1$ qubits randomly in four states, whose density is $I/2$,
 \begin{eqnarray}
 \rho_A={N+t_0\over N+t_0+t_1}{I\over 2}+{t_1\over N+t_0+t_1}{I \over 2}={I \over 2}.
 \end{eqnarray}
  The density matrix of qubit in  sequence $P_B$ is $I/2$. After the security check in step 3,  state $\ket{q_3}$ of Bob's qubit is "teleported" to Alice, in the form of $U^{-1}_T\ket{q_3}$. Alice, Bob and Charlie all know the explicit form of $U_T$ from the result of Bell-basis measurement. Alice and Charlie do not know $\ket{q_3}$. After Alice's encoding and Charlie's measurement, $U_m\ket{q_3}$ is announced publicly. Bob can derive $U_m$ by comparing $U_m\ket{q_3}$ with his initial state $\ket{q_3}$.
   Under noisy channel, after the security check, Alice could choose a classical linear code for the remaining qubits. The procedures are almost identical to those in the Shor-Preskill type of proof BB84 QKD protocol in Ref. \cite{shorpreskill}. The difference between QSDC and QKD is that the error correction encoding must be implemented before Alice sends her  sequence of $N+t_0$ encoded qubits to  Charlie. In the limit of large qubit numbers, the error rate threshold is 11\%.

   \paragraph{MDI-QSDC protocol using linear optical devices} Using linear optics, only two of the four Bell-basis states, $\ket{\psi^\pm}$, can be distinctively measured. Hence the MDI-QSDC protocol should be revised accordingly.
 Some modification to the 6 steps with full-Bell-basis measurements are described here. Step 1 is the same. In step 2, only $\ket{\psi^\pm}$ can be obtained after the Bell-basis measurements. When the measured result is $\ket{\phi^\pm}$, there is simply no clicks in the detectors. Therefore with high probability, only half of the qubits in $P_A$ and $P_B$ can give the results  $\ket{\psi^\pm}$ when  measured using linear optical device, as can be seen from Table \ref{t1}. There is no change in step 3, and eavesdropper will be found in the similar way as that in  the MDI-QKD. There is no change in step 4, the encoding operation is $U_m U_T$.  Steps 5 and 6 are also the same.

 \paragraph{Deterministic MDI-QKD protocol}  In a QKD protocol, eavesdropper can be found only after certain number of qubits have gone through the whole transmission process. Therefore when Eve is found, she has already acquired some transmitted data, which is disastrous for direct communicating secret message. In QSDC, block transmission is essential to prevent information leak before Eve's detection \cite{Long:2002ei}. For instance, in the security check in step 3, the comparisons of $t_1$ check pairs out of a block of $N+t_0+t_1$ pairs give a good estimate of the error rate, and ensures the security of qubits in $S_{Ah}$ before they are encoded with message and sent to Charlie. If the security of $S_{Ah}$ is not assured, Charlie or Eve could eavesdrop, for instance, Charlie does not make the Bell-basis measurement and stores the qubit in $P_A$ instead. After Alice encodes the message on her qubit in $S_{Ah}$ and sends it to Charlie, Charlie can perform Bell-basis measurement on the encoded qubit from $S_{Ah}$ and the stored qubit from $P_A$ to read Alice's encoded bit. Of course, Charlie's cheating will be found after dozens of rounds of transmission, the data transmitted before Alice and Bob finding her will be lost completely.

 Therefore the number of qubit in a block is reduced to 1, the QSDC protocol is no longer secure in sending secret message. Instead, it becomes a deterministic QKD protocol, namely it can still transmit random numbers deterministically and find eavesdropper after a session is finished. If an eavesdropper is found, they discard the transmitted data, otherwise they retain the transmitted data as raw key. The detailed procedure for the deterministic MDI-QKD is given below.

Step 1) Alice produces with probability $p_k=(N+t_0)/(N+t_0+t_1)$ an EPR pair in state $\ket{\psi^-_{12}}$, and with probability $p_c=1-p_k$ a single qubit (labeled qubit 2) randomly in one of the four states $\ket{+}$, $\ket{-}$, $\ket{0}$ and $\ket{1}$.  Meanwhile Bob prepares qubit 3  randomly in  one of the four states $\ket{+}$, $\ket{-}$, $\ket{0}$, $\ket{1}$.

Step 2).  Alice sends qubit 2, and Bob sends qubit 3 to Charlie.   Charlie performs Bell-basis measurement on the pair of qubits  and publishes the result.

Step 3)  After hearing from Charlie the Bell-basis measurement result,  Bob announces the basis of his qubit in the Bell-basis measured pair. Alice performs encoding operation $U_mU_T$ on qubit 1, where $m$ is a random bit and determined by Alice, $U_T$ is the unitary operation to complete the teleportation. Alice sends the encoded qubit 1 to Charlie, and Charlie first performs a unitary operation $U_B$ so that the qubit basis becomes $\ket{0}$, $\ket{1}$, namely $U_B=I$ if the basis of Bob's qubit is $\ket{0},\ket{1}$, and $U_B=H$  if  the basis of Bob's qubit is $\ket{+},\ket{-}$. Then Charlie measures qubit 1 in $\sigma_Z$ basis and publishes the result.

Step 4) After sufficient number $N+t_0+t_1$ of transmission has been performed ( $N+t_0$ number  of EPR-pairs and $t_1$ number of single qubits), Alice announces the initial states of the $t_1$ single qubits and Bob announces the initial states of corresponding single qubits, they will get an estimate of the error rate.  If the error rate is above a threshold, they conclude the transmission insecure and terminate the process. If the error rate is below a threshold, they conclude the transmission is secure. Then Alice announces the positions and bit values of the $t_0$ random check bits, Alice and Bob estimate the error rate. If the error rate is small, then they conclude the key distribution is  safe.  Alice and Bob will keep these $N$ random numbers as key.

 This MDI-QKD protocol does not use the block transmission technique, the security is  confirmed only after the distribution of the random numbers. This protocol could not send secret information directly, because all the transmitted data would leak to Eve before her detection. Classical communications are deterministic, but it cannot found eavesdropping. QKD can find eavesdropper, but cannot prevent Eve to access the transmitted data. QSDC can find eavesdropping and prevent Eve from obtaining the transmitted data.

 \paragraph{Discussion and Summary}We proposed a MDI-QSDC protocol using both EPR-pairs and single qubits. We also give a simplified version using linear optics devices, which only distinguishes two Bell-basis states. These MDI-QSDC protocols can be implemented with the present-day technology. Using the decoy method \cite{decoy1,decoy2} and the ILM-GLLP method  \cite{ILM,GLLP}, faint laser pulses and EPR-pairs from down conversion could be used with minor revisions to the protocols presented in this Letter. When the number of qubit in a block is reduced to 1, QSDC protocols reduce to deterministic QKD protocols.
 
\paragraph{Acknowledgement} This work was supported by the National Basic Research Program of China under Grant Nos. ~2017YFA0303700 and ~2015CB921001, National Natural Science Foundation of China under Grant Nos.~61726801, ~11474168 and ~11474181.

\bibliographystyle{apsrev4-1}
\bibliography{mdi-qsdc2}
\end{document}